# Effects of Sliding Speed on the Intensity of Triboluminescence in Slide contact: Experimental Measurements and Theoretical Analyses

Xuefeng Xu*

*School of Technology, Beijing Forestry University, Beijing 100083, China*

**Abstract**

Triboluminescence (TL) is the emission of light produced by rubbing or striking two materials together. Here, the light emission has been observed from the sliding contact between two disks under dry condition. The effects of the sliding speed on the intensity of TL have been experimentally investigated. The results show that the intensity of the emission light increases significantly with the sliding speed. A theoretical model is also proposed and an analytical expression is deduced for the intensity of TL in the slide contact. The theoretical prediction is found consistent with the experimental results. The present work may be helpful to the understanding of the mechanism of light emission when friction.

**Keywords:** Triboluminescence, Slide contact, Speed, Intensity.

---

* Corresponding author. E-mail: xuxuefeng@bjfu.edu.cn

# Ⅰ. Introduction

Triboluminescence (TL) is the emission of light when a material is pulled apart, scratched or crushed [1-4]. TL can occur in many everyday products such as sugar and salt. It has a wide range of application, including stress, fracture and damage sensors, display device, and X-ray photography. Although TL has been observed for several hundred years, its mechanism is still not well understood because there are too many factors that can affect TL.

Here, we focus our study on the effect of the sliding speed. First, experimental measurements are performed, and the variation of the intensity of TL with the slide speed is analyzed. Then, a theoretical model for the TL is presented, and an analytical formula for the TL intensity is deduced. Finally, the theoretical expression for the TL intensity is compared with the experimental results.

# Ⅱ. Experiments

The sliding friction experiments were performed by a disk-on-disk apparatus (see Fig. 1). In the experiments, a $SiO_2$ disk with a diameter of 30 mm rotates around its axis, and an $Al_2O_3$ disk with a diameter of 30 mm is pressed against the rotating disk by a normal force $F_N$. when friction, the emitted photons are collected by a quartz glass fiber and transferred to a spectrometer. A computer is used to record and analyze the light emission. By using this apparatus, two kinds of measurements can be done. One is the spectrum of the emitted light, and the other is the intensity of all the emitted photons with a wavelength from 200 nm to 900 nm.

# Ⅲ. Results and Discussion

## A. Spectrum of TL.

The measured spectrums of TL at different rotation speeds are shown in Fig. 2. The spectrums show strong and sharp peaks in the region of ultraviolet (UV), visible, and infrared (IR). It can be seen from the figure that the peaks in all the spectrums are the same. The coincidence of the spectrum peaks shows that the origin of the emitted light in the slide friction is the same for different relative slide speeds.

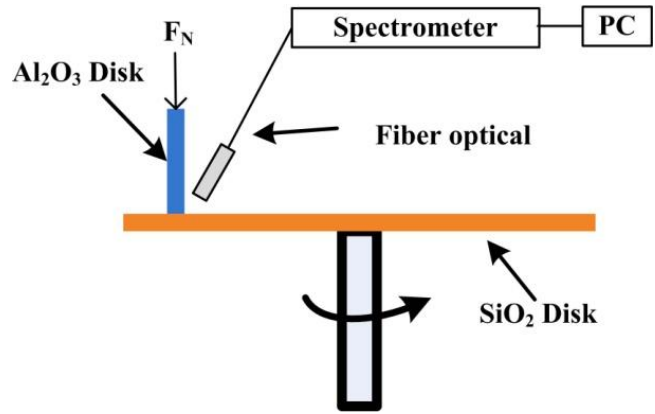

**FIG. 1.** The apparatus to measure the photons emitted from the sliding contact.

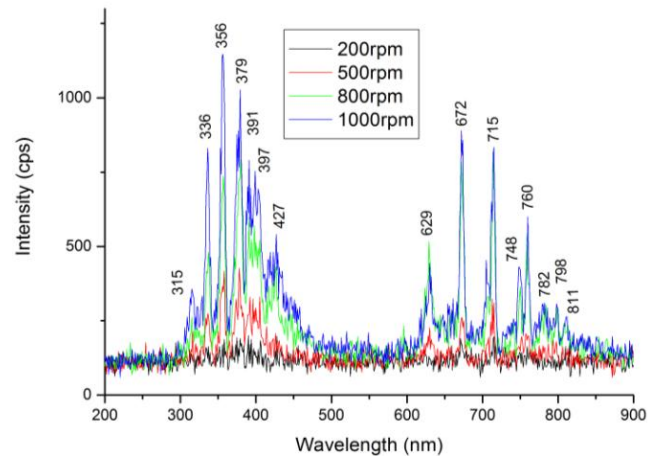

**FIG. 2.** Spectrums of TL at different rotation speeds.

As suggested by Nakayama [5], TL in the slide contact is generated by the discharging of gas molecules in the gap of the contact, which is in turn induced by the tribocharging on the mating surfaces (see Fig. 3). When friction, electrons will be transferred from a surface to the mating surface, resulting in the former potential being positive and the latter being negative, and thus forming an intense electric field in the gap of the contact. Under the action of the electric field, the electrons on the negatively charged surface will emit from the surface and accelerate in the gap. These moving electrons may collide with the air molecules in the gap. After the collision, the electrons inside the atoms of the molecules may be excited from the ground level to the exited levels. Then, photons will be emitted when the electrons in the excited level fall down to the lower or ground level. These emitted photons are observed as triboluminescence.

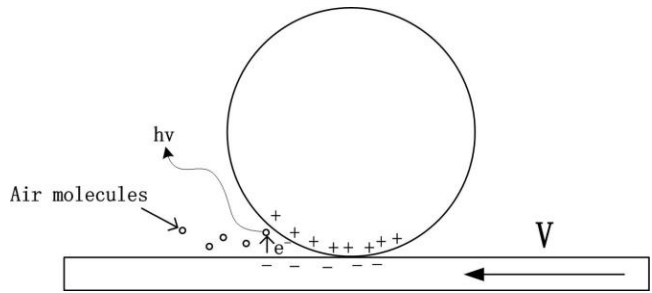

**FIG. 3.** Mechanism of TL generation.

**B. TL Intensity**

To reduce the time effect, we perform the measurement of TL intensity in a way like this: first, we do not rotate the disk and measure the noise signals for about 20 seconds; next, we turn on the machine and speed up to 200 rpm, and measure the light intensity of 200 rpm for about 20 seconds; then, we measure the light intensity of 400 rpm, 600 rpm, 800 rpm and 1000 rpm in turn in a similar way. This can ensure that a complete measurement including all the rotation speeds we considered can be performed in a time shorter than 150 seconds (see Fig. 4). After measurement, all the data of the same rotation speed will be averaged over the time first, and then will be subtracted by the time-averaged noise value.

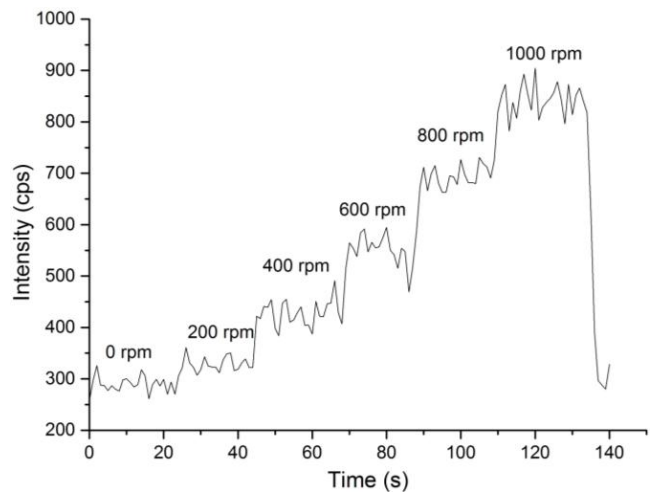

**FIG. 4.** The strategy of the measurement.

However, in different measurements, the light intensity may change a lot even if the rotation speed is the same (see Fig. 5). This may be attributed to the change in the relative position of the fiber and the contact point in different experiments, which will alter the collection efficiency of tribo-induced photon by the fiber. To remove this difference in measuring the TL intensity, all the data are normalized compared to the intensity at 1000rpm. After normalization, the data of all the experiments shows good consistency (see Fig. 6).

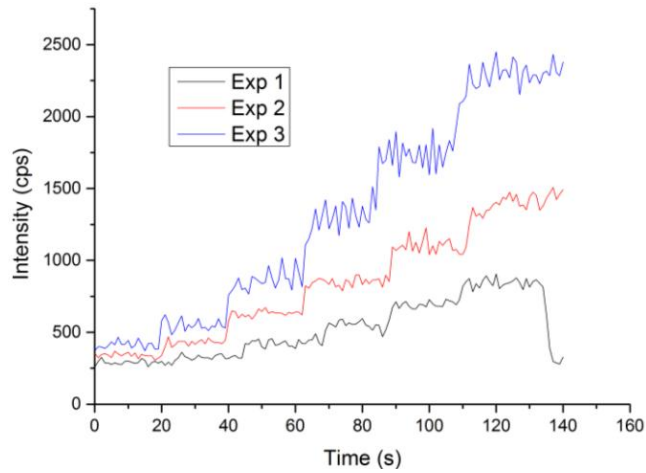

**FIG. 5.** The measured TL intensity in different experiments.

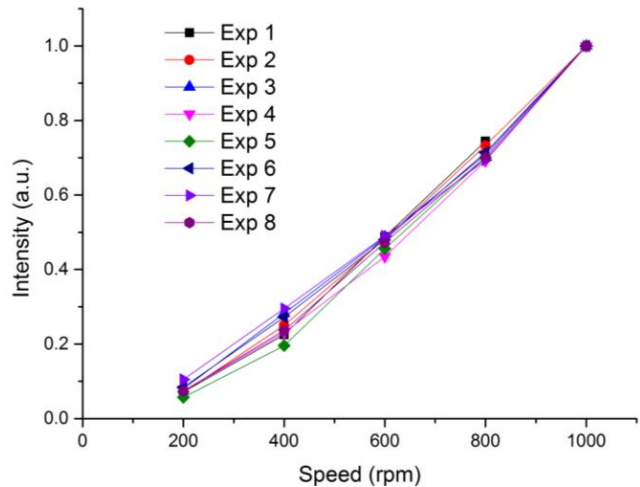

**FIG. 6.** The normalized data for TL intensity of different experiments. In the experiments, the normal for $F_N$=0.5N, the room temperature $T$=24$^o$, and the relative humidity $H$=45%.

Finally, the normalized data of the same speed in all the experiments are averaged, and a profile of the TL intensity as a function of sliding speed is obtained (see Fig. 7). It can be clearly seen that the TL intensity will increase with the sliding speed. In the following, a theoretical model for the TL intensity will be proposed to explain such an increase.

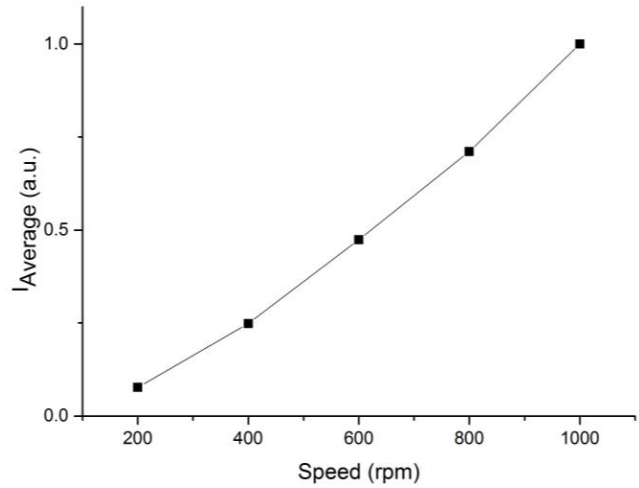

**FIG. 7.** The variation of TL intensity as a function of sliding speed.

**C. Theoretical Model for TL Intensity**

To obtain a model for the intensity of TL, the following assumptions are employed:

(1) Because the tribo-induced photon is generated by gas discharging which is in turn induced by the tribocharging, the photon intensity should be proportional to the number of the transferring electrons from one surface to the other.

(2) The energy needed for electron transfer is provided by the friction.

(3) The average energy rise of the surface electrons compared to their ground value should be proportional to the mechanical input energy, i.e., $\Delta \bar{E} = \alpha F_s V$, where $\Delta \bar{E}$ is the average energy rise of electrons, $\alpha$ is a coefficient which represents the averaged proportion of the energy transferred from the input mechanical energy to an excited surface electron, $F_s$ is the friction force, and $V$ is the line slide speed. When friction, $F_s$ do not change with the slide speed and thus can be considered as a constant in each measurement. We also assume that $\alpha$ is independent of the slide speed $V$ and the friction force $F_s$.

Then, according to the Boltzmann distribution law, in a group of electrons with an average energy rise of $\Delta \bar{E}$, the number of electrons with an energy rise of $\varepsilon$ can be expressed as

$$dN = Ce^{-\frac{\varepsilon}{\Delta \bar{E}}} d\varepsilon = Ce^{-\frac{\varepsilon}{\alpha F_s V}} d\varepsilon \qquad (1)$$

where $C$ is the normalization coefficient. Supposing that the energy rise need for electron to transfer is $\delta E$, the number of transferring electron equals to the number of surface electrons with an energy rise larger than $\delta E$ and can be computed as

$$N = \int dN = \int_{\delta E}^{\infty} Ce^{-\frac{\varepsilon}{\alpha F_s V}} d\varepsilon = C\alpha F_s V e^{-\frac{\delta E}{\alpha F_s} \cdot \frac{1}{V}} \qquad (2)$$

According to the assumption (1), the intensity of the friction-induced photons can be expressed as

$$I = \beta N = \beta C \alpha F_s V e^{-\frac{\delta E}{\alpha F_s} \cdot \frac{1}{V}} \qquad (3)$$

where $\beta$ is a coefficient which is independent of the slide speed $V$ and the friction force $F_s$. Taking logarithm of the equation (3), gives

$$Ln(I/V) = A - \frac{B}{V} \qquad (4)$$

where $A$ and $B$ are two positive coefficients.

To verify the validity of the present theoretical model, comparison between the theory and the experiments is performed. The Fig. 8 shows that the measured data of $Ln(I/V)$ have good linear dependence on $1/V$, which is consistent with the equation (4). This consistency may partly corroborate the validity of the present model for the TL intensity.

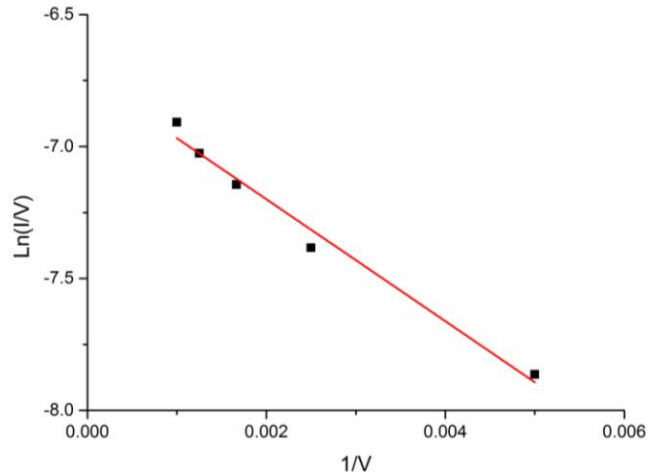

**FIG. 8.** Comparison between the theoretical model and the experimental results.

## Ⅳ. Conclusions

In the present work, an experimental apparatus was developed to observe the triboluminescence in the slide contact. By using this apparatus, we performed measurements on the intensity of the TL and found that the TL intensity increases as the sliding speed increases. Then, a theoretical model is proposed to compute the intensity of TL. By introducing some reasonable assumptions, a simple analytical expression for the TL intensity was deduced. The comparison between the analytical expression and the experimental measurements shows good agreement. The present work is useful to predict and control the intensity of TL during friction, and thus may be useful to understand the mechanism for TL generation.

## Acknowledgements

The work is financially supported by the National Natural Science Foundation of China (Grant NO. 51275050), and the Program for New Century Excellent Talents in University (Grant NO. NCET-12-0786).